\documentclass[a4paper,12pt]{article}
\usepackage[sumlimits,intlimits,namelimits]{amsmath}
\usepackage{amssymb}
\usepackage{a4wide}
\usepackage{xcolor}
\usepackage{multirow}
\usepackage{nicefrac}

\usepackage[english]{babel}
\makeatletter
\def\fmslash{\@ifnextchar[{\fmsl@sh}{\fmsl@sh[0mu]}}
\def\fmsl@sh[#1]#2{%
  \mathchoice
    {\@fmsl@sh\displaystyle{#1}{#2}}%
    {\@fmsl@sh\textstyle{#1}{#2}}%
    {\@fmsl@sh\scriptstyle{#1}{#2}}%
    {\@fmsl@sh\scriptscriptstyle{#1}{#2}}}
\def\@fmsl@sh#1#2#3{\m@th\ooalign{$\hfil#1\mkern#2/\hfil$\crcr$#1#3$}}
\makeatother
\definecolor{asparagus}{rgb}{0.53, 0.66, 0.42}
\usepackage{graphicx}
\usepackage{subfigure}
\usepackage{epstopdf}
\begin{document}
\begin{titlepage}
\begin{flushright}
SI-HEP-2023-01 \\ 
Nikhef-2023-001 \\[0.2cm]
\today
\end{flushright}

\vspace{1.2cm}
\begin{center}
{\Large\bf
Alternative Treatment of the Quark Mass in the \\[2mm] 
Heavy Quark Expansion}
\end{center}

\vspace{0.5cm}
\begin{center}
{\sc Anastasia Boushmelev, Thomas Mannel}   \\[2mm]
   Theoretische Physik 1, Center for Particle Physics Siegen \\
   Universit\"at Siegen,  D-57068 Siegen, Germany \\[5mm] 
{\sc K. Keri Vos} \\[2mm] 
{Gravitational Waves and Fundamental Physics (GWFP), \\
Maastricht University, Duboisdomein 30, NL-6229 GT Maastricht, the Netherlands} \\ and \\
{Nikhef, Science Park 105, NL-1098 XG Amsterdam, the Netherlands} 
\end{center}

\vspace{0.8cm}
\begin{abstract}
\vspace{0.2cm}\noindent
The treatment of the quark mass plays an important role when it comes to increasing the 
precision of the predictions of the heavy quark expansion for inclusive heavy hadron decays.
Various short-distance mass schemes have been invented to minimize the uncertainties induced by the quark mass, 
which needs to be extracted from other, independent observables.    
We suggest to replace the quark mass directly by an observable such as e.g. the inverse 
moments of the cross section for $e^+ e^- \to $ hadrons. We investigate this alternative 
strategy and study its impact on the perturbative series. 
\end{abstract}

\end{titlepage}

\newpage
\pagenumbering{arabic}
\section{Introduction}
Calculations in perturbative QCD have reached impressively high orders in the expansion in the strong 
coupling $\alpha_s$, opening new perspectives for the precision of theoretical predictions. However, 
since quarks and gluons are not the asymptotic states in QCD, one necessarily has to deal with   
the non-perturbative aspects of QCD to make predictions for observable quantities.  
Currently, in many cases, the non-perturbative part is the main factor limiting the precision 
of QCD predictions. 

In processes where a large scale $Q$ is present (such as a large momentum transfer or a large mass), 
many observables can be defined in terms of an Operator Product Expansion (OPE), which allows a 
factorization of short- from long-distance contributions. While the former can be computed in 
perturbation theory as a series in $\alpha_s(Q^2)$, the latter are parametrized 
in terms of hadronic matrix elements of operators expressed in terms of quark and 
gluon fields. These matrix elements have to be fixed by independent input, either 
from experiment or using non-perturbative methods such as lattice QCD. 

In many cases the leading term of the OPE is just the partonic, i.e. the perturbative 
contribution. It has been noticed long ago that the perturbative expansion of the short-distance
contributions is not a convergent series in $\alpha_s$, it can be at best an asymptotic series. 
This means that, starting at some order in the $\alpha_s$ expansion, the coefficients 
of the perturbative series start to grow, leading eventually to a divergent behaviour, 
although the first few terms look like a convergent series. In fact,  
certain divergent contributions have been identified by summing the leading terms 
of a (formal) large $n_f$, the number of active quark flavours, expansion, 
which are diagrammatically presented by
``bubble chains''. These diagrams lead to a factorial divergence which is 
still Borel summable, but the Borel transform exhibits poles on the positive real 
axis, the so called infrared renormalons (for a review on this see \cite{Beneke:1998ui}).  

These infrared renormalons lead to ambiguities in the expressions for the observables, 
which relate to the power suppressed terms in the OPE and also (in the cases we shall
discuss) to the definition of the quark mass. In turn, neither the quark mass nor matrix elements appearing in the power-suppressed terms of the OPE are physical quantities, since they need to be defined.  

In heavy-quark physics, the large scale is set by the mass of the heavy quark, consequently 
the OPE is a series in inverse  powers of the heavy-quark mass. The starting point of any 
perturbative calculation is usually the  pole mass for the quarks, which, however, is a 
purely perturbative concept. Its advantage is that it is gauge and scale independent. However, 
its disadvantage is that it suffers from infrared renormalons
\cite{Bigi:1994em,Beneke:1994sw,Neubert:1994wq}, leading to  
ambiguities of order $\Lambda_{\rm QCD}$ in the definition of the heavy quark mass.  
Strictly speaking, this would make a systematic heavy-quark expansion in this parameter impossible.

To this end, many different mass definitions have been invented, tailored to the specific 
application. At high energies, the $\overline{\rm MS}$ mass is often used, but this is 
restricted to scales $\mu \ge m$ \cite{Chetyrkin:2010ic}. 
In heavy-quark physics, a mass definition that 
can be used at scales $\mu \le m$ is needed, such as the kinetic mass 
\cite{Bigi:1994ga,Fael:2020iea,Fael:2020njb} 
or the $1S$ mass \cite{Hoang:1998ng,Hoang:1998hm}. These mass definitions solve the problem of the renormalon ambiguity, and can be determined 
quite precisely from other independent sources. 

Although both mass schemes are well established in $B$-meson decays, transferring similar mass definitions to the lighter $D$-meson decays is more challenging \cite{Gambino:2010jz,Fael:2019umf}. 
In the kinetic mass scheme, a hard cut-off scale $\mu$ is introduced, which has to be 
perturbative, i.e. $\alpha_s (\mu) < 1$. However, it should also satisfy $\mu / m_Q < 1$. 
While this can be achieved for the $b$ quark by choosing $\mu \sim 1$ GeV, there is 
no window of this kind for the charm quark. Likewise, in the $1S$ scheme, one assumes 
that the $1S$ state of a $Q \bar{Q}$ quarkonium can be treated as a Coulombic system, i.e. 
can be treated perturbatively. While this seems to be valid for the bottomonium system, 
this is clearly not true for the corresponding $1S$ charmonium.

The usual strategy for practical calculations is to define the quark masses as well 
as the matrix elements appearing as power corrections in the OPEs of observables within  
a specific scheme and to obtain values for these parameters from fits to data. However, 
these parameters do not have any physical meaning, so one may also take the point of 
view that these parameters only transport the information obtained form one observable  
another observable. This suggests a slightly different 
strategy for perturbative QCD calculations in the framework of OPE. 

It is generally assumed that the ambiguities induced by infrared renormalons cancel 
between the perturbative series and properly defined non-perturbative quantities, 
including also the quark mass, such that observables are free of such ambiguities. 
In fact, this has been shown explicitly in the bubble chain approximation for both 
the HQE and the $e^+ e^-$ moments \cite{Beneke:1994sw,Neubert:1994wq}. In order to fix these ambiguities, one has to chose for each parameter an additional observable, which is 
also computed in terms of an OPE up to the desired order in the $\alpha_s$ and the power expansion. 

Usually these parameters are extracted explicitly and used as general parameters. However, 
one may as well proceed in an alternative way and solve the expression obtained from 
the OPEs for the parameters and eliminate the parameters from the expressions for 
the observables. 

In this way, one constructs  
perturbative relations between observables without any reference to quark masses or 
the parameters appearing in the power corrections of the OPE. 
We note that the OPEs for the observables has to be truncated at some order of the power 
expansion, which implies that the perturbative relations between the observables 
will still suffer from divergences, which are due to renormalon ambiguities induced by  
higher orders in the OPE \cite{Martinelli:1995zw}. However, one would expect that the 
divergent behaviour of the perturbative series is shifted to higher and higher orders in $\alpha_s$, 
the more power corrections are included and fixed by new observables.

In the present paper, we shall discuss this new treatment quantitatively for the Heavy Quark Expansion (HQE) and the inverse moments of the $e^+ e^- \to $ (heavy) hadron cross section focused on eliminating the heavy quark mass.   
In the next section, we will give more details on the general strategy and     
set the stage for the discussion of the role of the heavy quark mass in the HQE. 
In Sec.~\ref{sec:three}, we first discuss the moments of the cross section 
for $e^+ e^- \to $ heavy hadrons, construct the perturbative relations between 
different inverse moments and study the resulting series in $\alpha_s$.  
We then apply the same strategy to the inclusive $B\to X_u \ell \bar\nu$ decay rate, 
relating it to the inverse moments of the cross section for 
$e^+ e^- \to $ heavy hadrons and discuss the resulting perturbative series.

\section{Infrared Renormalons in a Nutshell} \label{sec:one} 
In this section, we outline the strategy we pursue for the calculations.  
We start from an observable $M$ which has an OPE in inverse powers of the heavy quark 
mass  
\begin{equation} \label{MasterOPE}
M(m_Q) = \sum_n \sum_i C_n^{(i)}(m_Q) \langle  O_n^{(i)} \rangle  \qquad C_n^{(i)} \sim \frac{1}{m_Q^n}
\end{equation}
where $O_n^{(i)}$ are operators of mass-dimension $n$ and $\langle ... \rangle$ denotes a forward matrix element with appropriate states. The coefficients $C_n^{(i)}(Q)$ 
can be computed in perturbation theory as a series in $\alpha_s (Q)$, while the matrix elements 
$\langle  O_n^{(i)} \rangle \sim \Lambda_{\rm QCD}^n$ encode the non-perturbative input.

The coefficients $C_n^{(i)}(m_Q)$ in the OPE can be expanded in powers of $\alpha_s (m_Q)$, and
the running of $\alpha_s (\mu)$ is governed 
by the $\beta$ function, which can be chosen such that it depends on $\alpha_s$ only:
\begin{equation}
	- \frac{1}{2}  \frac{d \ln  \alpha_s (\mu)}{d \ln \mu} = \beta(\alpha_s (\mu)), \quad \beta (\alpha_s) = \frac{\alpha_s}{\pi } \beta_0 + \cdots, \quad \beta_0 = \left(\frac{11}{12} C_A - \frac{1}{3} T_f n_f\right) 
\end{equation}
where $n_f$ is the number of active quark flavours and $C_F=4/3$, $C_A = 3 $ and $T_f = 1/2$ for $SU(3)$. The solution of this equation can be written as  
\begin{equation}
\ln \left(\frac{\mu^2}{m_Q^2}\right) =  \int\limits_{\alpha_s(\mu)}^{\alpha_s(m_Q)} \frac{d \tau}{\tau} \frac{1}{\beta(\tau)}, \qquad 
\frac{\mu^2}{m_Q^2} = \exp \left(\int\limits_{\alpha_s(\mu)}^{\alpha_s(m_Q)} \frac{d \tau}{\tau} \frac{1}{\beta(\tau)} \right) 
\end{equation} 
In the following we will concentrate on the leading term in the $\beta$ function, which implies
\begin{equation}
\int\limits_{\alpha_s(\mu)}^{\alpha_s(m_Q)} \frac{d \tau}{\tau} \frac{1}{\beta(\tau)} 
= \frac{ \pi}{\beta_0} \int\limits_{\alpha_s(\mu)}^{\alpha_s(m_Q)} \frac{d \tau}{\tau^2}  
= \frac{\pi}{\beta_0} \left( \frac{1}{\alpha_s(\mu)} - \frac{1}{\alpha_s (m_Q)}  \right)  
= \ln \left( \frac{\mu^2}{m_Q^2} \right)
\end{equation} 

Furthermore, we define a scale parameter $\Lambda_{\rm QCD}$ by 
\begin{equation}
\Lambda_{\rm QCD}^2 = m_Q^2  
\exp \left(\int\limits_{\infty}^{\alpha_s(m_Q)} \frac{d \tau}{\tau} \frac{1}{\beta(\tau)} \right)  
 \approx m_Q^2 \exp \left(- \frac{ \pi}{\beta_0} \frac{1}{\alpha_s (m_Q)} \right) 
\end{equation} 

For the cases we shall consider the first term of the OPE observable $M$
is given by the perturbative expression 
\begin{equation}\label{eq:mpert}
	M_{\rm pert} (m_Q) = (m_Q)^d C_0 (\alpha_s(m_Q)) = (m_Q)^d \sum_{n=0}^\infty C_0^{(n)} 
 \left(\frac{\alpha_s (m_Q)}{\pi} \right)^n 
\end{equation}
where $a_n$ are the coefficients of the perturbative expansion and $d$ is the mass dimension of the 
observable. Note that the dimensionality of $M_{\rm pert}$ has to come from the quark mass, because there is no other scale involved. 

However, it is well known and established, that the series in $\alpha_s (m_Q)$ is at best an 
asymptotic series. This conjecture is supported by calculations in the large-$\beta_0$ limit.  
The perturbative series exhibits contributions with factorially growing coefficients of the form 
\begin{equation} \label{coeffak} 
a_k \sim  \left(   \beta_0   \right)^k \, k! \, b_k
\end{equation}  
with $b_k \sim 1$, indicating a problem with the perturbative series. 

The divergent behaviour of the perturbative series is closely related to the OPE for the observable. Within the OPE, we obtain power-suppressed  
contributions of the from $(\Lambda_{\rm QCD} / Q)^n$. Re-expressing this expression in terms
of $\alpha_s$, we get 
\begin{equation}
	(\Lambda_{\rm QCD} / m_Q)^n =  \exp \left(- \frac{n}{2} \frac{\pi}{\beta_0} \frac{1}{\alpha_s (m_Q)} \right)
\end{equation} 
which clearly does not have a Taylor series in $\alpha_s (m_Q)$. 

To clarify the relation of the divergent behaviour of the perturbative series and the power like contribution 
obtained form the OPE, we look at the Borel transform $S(u)$ of a function 
$F(x)$ defined by its (formal) Taylor series 
\begin{equation}
F(x)	= \sum_{k=0}^\infty a_k x^k  \qquad \stackrel{\rm Borel-Trf}{\longrightarrow}  \qquad  
S(u) = \sum_{k=0}^\infty \frac{a_k}{k!} u^k  
\end{equation} 
In case the taylor series for $C(x)$ is convergent for positive values of $x$, we can reconstruct $F(x)$ from $S(u)$ by computing the integral
\begin{equation} \label{BT}
F(x) = \int\limits_0^\infty e^{-t} S (x t) \,  dt
\end{equation}
However, the presence of the factorially growing contributions prevent us from computing the 
integral. In fact, assuming $a_k = k!$, we get 
\begin{equation}
F(x)	= \sum_{k=0}^\infty k! x^k  \qquad \stackrel{\rm Borel-Trf}{\longrightarrow}  \qquad  
S(u) = \sum_{k=0}^\infty  u^k = \frac{1}{1-u} \ .
\end{equation} 
This pole occurring in $S(u)$ at positive $u$ prevents us from performing the $t$-integral in \eqref{BT}. Thus one needs to give a prescription of how to deal with the pole, which introduces an ambiguity in the inverse transformation (\ref{BT}). One way to proceed is to deform the integral in (\ref{BT}) into a contour integral in the complex t plane, and the ambiguity can be defined as the difference between circling the pole by moving into the upper $t$-half plane and by moving to the lower $t$-half plane. This leads to a ``localized''
ambiguity of the form
\begin{equation}
    \Delta S = \eta \, \delta(1-u)  
\end{equation}
which translate to a corresponding ambiguity in $C(x)$ 
\begin{equation}
    \Delta F (x) = \eta \exp\left(- \frac{1}{x} \right)
\end{equation}

We may apply this to the perturbative series discussed above. To make the argument clear, we use (\ref{coeffak}) and set $b_k = \eta$ (i.e. $b_k$ independent of $k$). Arguing along the lines described 
above, this leads to an ambiguity in the observable of the form
\begin{equation}
    \Delta M = \eta \, \exp( -   \frac{\pi }{\alpha_s (m_Q)\beta_0} ) = \eta \, \left(\frac{\Lambda_{\rm QCD}}{m_Q}\right)^2
\end{equation}
encoded in a power suppressed term. 
 
In the general case, multiple poles and even other types of singularities appear, hindering us to compute the integral (\ref{BT}) without additional prescriptions how to deal with the 
singularities. 

In addition to this, it is well known that also the quark mass can introduce a renormalon problem. Since the quark mass 
is not a physical parameter, one needs to chose a mass scheme for its definition. The staring point of perturbative
calculations is usually the pole mass $m_Q^{\rm pole}$, defined by the position of the pole in a perturbative 
calculation. This prescription provides a recipe to remove the ultraviolet divergence from the quark self energy, 
and any renormalized mass $m_Q^{\rm ren}$ defined in any other scheme can be related to the pole mass by a perturbative 
calculation with finite coefficients $r_i$ 
\begin{equation} \label{masrel}
    m_Q^{\rm pole} = m_Q^{\rm ren} \left(1+
    \sum_{i=1}^\infty r_i \left( \frac{\alpha_s  (m_Q^{\rm ren})}{\pi} 
    \right)^{\!i} \right) \ .
\end{equation}
This perturbative relation suffers from renormalon ambiguities as soon as $m_Q^{\rm ren}$ is a so called 
short-distance mass. To be explicit, we quote the relation between the mass $m_Q^{\overline{\rm MS}}$ defined in the
$\overline{\rm MS}$ scheme and the pole mass, calculated in the ``bubble chain'' approximation. The Borel transform 
of the perturbative series in \eqref{masrel} exhibits poles \cite{Neubert:1994wq} 
\begin{eqnarray}
    m_Q^{\rm pole} (u) &=& m_Q^{\overline{\rm MS} } (m_Q) \frac{C_F}{\beta_0}  \\ \nonumber 
    && \times \left(- 2e^{5/6}  \frac{1}{u-\frac{1}{2}} + \frac{3}{4}e^{5/2} 
    \frac{1}{u-\frac{3}{2}} 
    - \frac{1}{2}e^{10/3} \frac{1}{u-2} + \frac{9}{64}e^{25/6} \frac{1}{u-\frac{5}{2}} + \cdots  \right) 
\end{eqnarray}
inducing in general ambiguities of the form  
\begin{equation}
    m_Q^{\rm pole} (u) = m_Q^{\rm ren} \left( 1 +  \sum_{n=1}^\infty \eta_n \left(\frac{\Lambda_{\rm QCD}}{m_Q} \right)^n  \right)  \ ,
\end{equation} 
where $\eta$ is the residue of the pole at $u_n = 1/2, 3/2, 2, ..$
in the relation of the pole mass to some short-distance mass. Note that the term of order $(\Lambda_{\rm QCD}/m_Q)^2$ 
is absent in the relation between the pole mass and the $\overline{\rm MS}$ mass in this bubble-chain approximation.

It is generally assumed and explicitly demonstrated in the bubble-chain approximation that the renormalon 
ambiguities have to cancel in physical observables \cite{Beneke:1994sw,Neubert:1994wq,Martinelli:1996pk}. 
However, the perturbative series remains asymptotic, even if some of the renormalon ambiguities are 
resoved \cite{Martinelli:1996pk}, so there is no possibility to fix the perturbative series without 
resumming all power corrections. However, the onset of the asymptotic behaviour can be delayed by 
removing renormalon ambiguities. 

In particular, the renormalon in the pole mass cancels against (some of) the divergent 
behaviour in the perturbative expansions of the Wilson coefficients in the OPE. Therefore, we expect 
that the choice of a renormalon-free short distance mass will shift the onset of the 
divergent behaviour of the perturbative series to higher orders.   
Likewise, a proper definition of the matrix elements $\langle  O_n^{(i)} \rangle$ will remove 
further renormalon ambiguities and will further delay the onset of the divergent behaviour of 
perturbation theory. 

In practice, the quark mass (in any scheme) as well as the matrix elements $\langle  O_n^{(i)} \rangle$ 
are obtained from other, independent observables, which we assume also to have an OPE. To this end, for a calculation 
up to some power $(\Lambda_{\rm QCD} / m_Q)^k$ in the OPE we will have - aside from the quark mass - a finite number 
$j$ of matrix elements $\langle  O_n^{(i)} \rangle$, since $n \le k$. 
These can be determined from $j+1$ observables that have an OPE like (\ref{MasterOPE}), 
which is expanded to the same order $k$. In this way, all unknown parameters, including the quark mass, are fixed in terms of observables. In the following we pursue this strategy, namely removing the quark mass(es) and the hadronic matrix elements by inserting observables. 

Doing so, generates a relation between observables expressed in terms of a perturbative expansion. As stated above, 
this perturbative expansion is still plagued by divergences, which are related to renormalon ambiguities induced by 
terms of higher order $(\Lambda_{\rm QCD}/m_Q)^{k+1}$ in the OPEs of the observables. In terms of the Borel transform 
of the perturbative contribution this means that ambiguities related to the first $k$ poles have been removed, leaving poles located at $u \ge \frac{k+1}{2}$. The lowest pole at $u = \frac{k+1}{2}$  contributes to the factorial 
divergences as
\begin{equation}
C_0^{(n)} \sim  \left( \frac{ \beta_0}{4 }  \right)^n \, n! \left(\frac{1}{k+1} \right)^{n+1}
\end{equation}
showing the suppression factor $1/(k+1)^{n+1}$ which eventually leads to a later onset of the asymptotic 
behaviour of perturbation theory. 

In the following we test this conjecture by explicit calculation, using the inverse moments of the 
$e^+ e^- \to $ hadrons cross section and the heavy quark expansion for semileptonic decays as an example.

\section{Replacing the Quark Mass through Observables}\label{sec:three}
\subsection{$e^+e^-$ Moments}
Historically, the cross section for $e^+e^- \to $ hadrons has been the prime quantity to discuss the notion of quark-hadron duality. The underlying elementary process is $e^+e^- \to $ quarks and gluons, and the early understanding of duality was that the partonic cross section $e^+e^- \to $ quarks and gluons should be identical to the $e^+e^- \to $ hadrons, once an appropriate ``smearing'' is applied. This idea has been quantified by considering moments of the cross section, and the statement of duality turns into the statement that the moments of the cross section $e^+e^- \to $ hadrons should be the same within the precision of the calculation of $e^+e^- \to $ quarks and gluons. 

In a more modern language, the notion of duality is linked to the validity of an OPE for the relevant observable. For the case at hand, we exploit the relation between the vacuum polarization and the ratio 
\begin{equation}
    R(s) = \frac{\sigma(e^+ e^- \to {\rm hadrons})}{\sigma(e^+ e^- \to \mu^+ \mu^-)} \ .
\end{equation}
The dispersion relation between of the vacuum polarization contains the $R$ ratio 
\begin{equation} \label{DispRel}
    \Pi(q^2)=\frac{1}{12 \pi}\int \frac{\text{d}s}{s}\dfrac{R(s)}{s-q^2} \, , 
\end{equation}
where the vacuum polarization function is
\begin{equation}
    \Pi_{\mu\nu}(q^2)=i\int \text{d}^4x\exp(iqx)
\langle 0|T [\mathcal{J}_{\mu}(0)\mathcal{J}_{\nu}(0)]|0\rangle=(q_{\mu}q_{\nu}-q^2g_{\mu\nu})
    \Pi(q^2)
\end{equation}
where 
$\mathcal{J}_{\mu}(0)=\sum_f Q_f\overline{q}_f\gamma_{\mu}q_f$ is the electromagnetic current, $Q_f$ is the charge of the quark $f$ and $q$ is the momentum transfer with $s=q^2$.

For sufficiently large $s$ the quantity $\Pi (s)$ has an OPE of the form 
\begin{equation} \label{PiOPE} 
 \Pi(s) = \sum_{n=0}^\infty \sum_i  
  a_i^{(n)} (s) \langle 0 | O_i^{(n)} | 0 \rangle  
\end{equation}
where $O_i^{(n)}$ are a set of local operators (labelled by $i$) of dimension $n$ and 
$a_i^{(n)} (s)$ are coefficients calculable in perturbation theory. As discussed previously, the leading term in (\ref{PiOPE}), for $n=0$, is just the perturbative 
contribution. The first non-vanishing non-perturbative contributions appear at dimension four and involve the quark and gluon condensates.  

To relate hadronic and partonic quantities, we define the moments $R(s)$ according to 
\begin{equation}
    M_n = \int \frac{ds}{s}  \frac{1}{s^n} R(s) 
\end{equation} 
Note that $R(s)$ is dimensionless and approaches a constant as $s \to \infty$, so only 
moments with $n \ge 1$ can be computed.  

Inserting the dispersion relation (\ref{DispRel}) we can relate these moments to  
$\Pi(s)$ 
\begin{equation} \label{MnvsPi} 
    M_n=\frac{12 \pi^2}{n!}\left(\frac{\text{d}}{\text{d}q^2}\right)^{\!n}\Pi(q^2)\bigg \vert_{q^2=0} \, , 
\end{equation}
and, combining this with (\ref{PiOPE}), shows that one obtains an OPE for $M_n$ as well. 
This observation is the exact formulation of the relation between ``smeared'' $e^+ e^-$ 
cross sections and the formulation of duality in terms of the OPE. 

We are especially interested in the contribution $R_Q$ from the heavy charm and bottom quarks. In this case, the partonic cross section vanishes below the 
threshold value 
$s_{\rm th} = 4 m_Q^2$, i.e. 
$R_Q(s) = 0$ for $s \le s_{\rm th}$.  
In \cite{Chetyrkin:1996cf} the leading term, i.e. the partonic part has been computed in a
perturbative series in $\alpha_s$. In particular, the Taylor series of $\Pi(s)$ reads 
\begin{equation}\label{eq:Cienter}
    \Pi(q^2)^{\rm pert}=\frac{3}{16\pi^2} Q_Q^2\sum_{n>0} C_n z^n
\end{equation}
with $z=q^2/(4m_Q^2)$ and $Q_Q$ the charge of the heavy quark. 
The coefficients $C_n$ 
\begin{equation}
 C_n= \sum_k C_n^{(k)} \left(\frac{\alpha_s}{\pi}\right)^{\!k} \ ,
\end{equation} 
are computed perturbatively using the pole mass to order $k=2$ in \cite{Chetyrkin:1996cf,Chetyrkin:1997mb}. The four-loop contributions of $C_0$ and $C_1$ were computed in \cite{Boughezal:2006px, Chetyrkin:2006xg}. To be concrete, the coefficients  
$C_n^{(k)}$ to each order contain logarithms of the form $\ln (\mu^2/m_Q^{2})$, 
and the series reads 
\begin{eqnarray}
C_n &=& C_n^{(0)} + \sum_{i=1}^\infty  \sum_{k=0}^{i-1} 
\left(\frac{\alpha_s(\mu)}{\pi}\right)^{\!i} C_n^{(i,k)} 
\left( \ln\left(\frac{\mu^2}{m_Q^2}\right)\right)^{\!k}  
\\ \nonumber &=& C_n^{(0)}+C_n^{(1)}\left(\frac{\alpha_s(\mu)}{\pi}\right)+\left(C_n^{(2,0)}+C_n^{(2,1)}\ln\left(\frac{\mu^2}{m_Q^2}\right)\right)\left(\frac{\alpha_s(\mu)}{\pi}\right)^{\!2}+ \ldots \, .
\end{eqnarray}
where $C_n^{(1)} \equiv C_n^{(1,0)}$. 
Note that we write explicitly the $\mu$ dependence, which is spurious, since 
the $C_n$ overall do not depend on $\mu$. In other words, the explicit $\mu$ dependence 
from the logarithms has to cancel the one of $\alpha_s$ order by order. This means 
that, for the cases at hand, we have 
\begin{equation}\label{eq:cnbeta0}
C_n^{(2,1)} = \beta_0 C_n^{(1)} 
\end{equation}
such that the $\mu$ dependence is in fact ${\cal O} (\alpha_s^3)$. 

According to (\ref{MnvsPi}) the 
inverse moments are given in terms of the coefficients of the Taylor series of $\Pi (q^2)$ 
as 
\begin{equation}  
    M_n^{\rm pert} = \frac{9}{4} Q_Q^2 \left(\frac{1}{2m_Q^{\rm pole}}\right)^{\!2n} \sum_k C_n^{(k)} \left(\frac{\alpha_s}{\pi}\right)^k =\frac{9}{4} Q_Q^2 \left(\frac{1}{2m_Q^{\rm pole}} \right)^{\!2n}  F_n (\alpha_s (\mu)) \, ,
    \label{eq:mom-mass}
\end{equation}
which is used e.g.\ to extract the quark masses from the 
inverse moments of the $e^+ e^- \to$ hadrons cross sections \cite{Kuhn:2007vp}.

Power-suppressed contributions to $\Pi$ appear at (relative) order $(\Lambda_{\rm QCD}/m_Q)^4$ and are given in terms of 
quark- and gluon condensates \cite{Shifman:1978bx}. For heavy quarks, the quark condensate is absent, therefore only the gluon condensate contributes. For the moments, this gives \cite{Shifman:1978bx,Shifman:1978by}  
\begin{equation}
     M_n^{\langle GG \rangle} =\frac{12 \pi^2 Q_Q^2}{(2m_Q^{\rm pole})^{2n}}
     \frac{
     \left\langle \frac{\alpha_s}{\pi}G_{\mu \nu}^a G^{a,\mu\nu} \right\rangle}{(2 m_Q^{\rm pole})^4}
     \, a_n \left(1+b_n  \frac{\alpha_s}{\pi}\right)
\end{equation}
where the coefficients $a_n$ and $b_n$ can be found in \cite{Broadhurst:1994qj}.    

The value of the gluon condensate is not very precisely known, but it is 
small and even compatible with zero \cite{Broadhurst:1994qj,Colangelo:2000dp}. 
For this reason it is often neglected, at 
least for moments with small $n$. This   
means that for all practical calculations the perturbative contribution, i.e. the leading term of the OPE is 
sufficient, and consequently we drop the contributions from the gluon condensate.     

Following the strategy outlined in the last section, we consider at the Borel transform of the perturbative
series of the leading term. Schematically, we expect it to have the form 
\begin{equation} \label{RenPole} 
 F_n (\alpha_s (\mu)) \to S_n (u) = \frac{R_{1/2}}{u-\frac{1}{2}} 
     + \frac{R_1}{u-1}  +
    \frac{R_{3/2}}{u-\frac{3}{2}} + \cdots 
\end{equation}
where the ellipses denote terms related to renormalon singularities at even higher values of $u$.
Note that we have dropped the first power-suppressed terms, which are related to poles in (\ref{RenPole}) 
at $u \ge 2$ so the renormalon poles shown explicitly in (\ref{RenPole}) are related to the pole mass. 

The usual way to deal with this is to define a renormlized short distance mass $m_Q^{\rm ren}$ defined in 
(\ref{masrel}). In fact, we may insert the expression (\ref{masrel}) 
of the pole mass in terms of a short distance mass into \eqref{eq:mom-mass} and re-expand into the
perturbative series\footnote{Note that $F_n ( \alpha_s (\mu)) $ in \eqref{eq:mom-mass} does not depend on $\mu$, so we can set $\mu = m_Q^{\rm ren}$. 
} 
\begin{eqnarray}
     M_n^{\rm pert} &=& \frac{9}{4}Q_Q^2 \left(\frac{1}{2 m_Q^{\rm ren}} \right)^{\!2n}   \left(1+\sum_{i=1}^\infty r_i \left(\frac{\alpha_s (m_Q^{\rm ren})}{\pi} \right)^{\! \! i} \,  \right)^{\!-2n} F_n (\alpha_s (m_Q^{\rm ren}))  \\ \nonumber 
     &=& \frac{9}{4} Q_Q^2 \left(\frac{1}{2 m_Q^{\rm ren}} \right)^{\!2n}  
     \left(C_n^{(0)} +   
     \left(\frac{\alpha_s (\mu)}{\pi}\right) ( C_n^{(1)} - 2 n r_1 C_n^{(0)}) \right. 
     \\ \nonumber 
     && \left. + \left(\frac{\alpha_s (\mu)}{\pi}\right)^{\!2} ( C_n^{(2)} - 2 n r_1 C_n^{(1)} + 
     n C_n^{(0)} [(2n+1)r_1^2 - 2 r_2]) + \cdots 
     \right) 
\end{eqnarray}
Assuming the cancellation of the ambiguities related to the quark mass, the Borel transform of the 
resulting perturbative series does not have poles at $u=1/2,\, 1$ and $3/2$, since power corrections, 
i.e. the gluon condensate, appear only at $1/m^4$, related to poles at $u=2$ and higher. 
We note in passing that the coefficients of the resulting series depend explicitly on $n$. 
This indicates a problem since large values of $n$ may jeopardize the perturbative series. We 
shall return to this point below. 

According to the strategy outlined above, we may as well replace the pole mass by an observable, 
e.g. by another inverse moment. Solving one of the inverse moments for the pole mass  
\begin{equation}\label{eq:massinmom}
m_Q = \frac{1}{2}\left(\frac{9}{4}Q_Q^2\right)^{1/(2n)} \left(\frac{C_n}{M_n}\right)^{1/(2n)}  
\end{equation}
we obtain an expression that can be inserted into other moments. By the same argument as above the 
resulting perturbative series for the relation between two inverse moments will not have 
renormalon ambiguities for $u=1/2,\, 1$ and $3/2$. Therefore, we expect that this results in a better behaviour of the perturbative series.

\begin{table}[t]
\centering
\begin{tabular}{c|ccccccc}
$k$&\multicolumn{7}{ c }{$n$} \\
&	1	&	2	&	3	&	4	&	5	&	6	&	7	\\ \hline
$a_{1,n}^{(1)}$& & 1.18 & 1.76 & 2.12 & 2.38 & 2.58 & 2.73 \\
$a_{1,n}^{(2)}$& & -1.27 & -0.30 & 0.84 & 1.90 & 2.84 &3.67 \\ \hline 
$a_{2,n}^{(1)}$& -2.36 &  & 1.15 & 1.88 & 2.40 & 2.79 & 3.11 \\
$a_{2,n}^{(2)}$& 6.73 &  & 0.92 & 2.89 & 4.94 & 6.87 & 8.63 \\ \hline
$a_{3,n}^{(1)}$& -5.27 & -1.72 & & 1.09 & 1.87 & 2.46 & 2.94 \\
$a_{3,n}^{(2)}$& 19.4 & 1.09 & & 1.89 & 4.46 & 7.10 & 9.63 \\ \hline
$a_{4,n}^{(1)}$&-8.48 & -3.75 & -1.46 &  & 1.04 & 1.83 & 2.46 \\
$a_{4,n}^{(2)}$&41.6 & 4.80 & -0.66 &  & 2.42 & 5.37 &8.39 \\ \hline
$a_{5,n}^{(1)}$& -11.9 & -5.99 & -3.12 & -1.30 & & 0.99 & 1.77 \\
$a_{5,n}^{(2)}$& 75.5 & 12.8 & 0.35 & -1.52 & & 2.75 & 5.93 \\ \hline
$a_{6,n}^{(1)}$& -15.5 & -8.37 & -4.93 & -2.74 & -1.19 & & 0.94 \\
$a_{6,n}^{(2)}$& 122. & 26.1 & 4.01 & -1.79 & -2.01 & & 2.96 \\ \hline
$a_{7,n}^{(1)}$& -19.1 & -10.9 &-6.85 & -4.30 & -2.48 & -1.10 & \\
$a_{7,n}^{(2)}$& 184. & 45.8 &11.0 & -0.16 & -3.01 & -2.33 & \\
\end{tabular}
\caption{Coefficients $a^{(1)}_{k,n}$ and $a^{(2)}_{k,n}$ for different values of $n$ and $k$ using $n_l=4$. 
}
\label{table:coeff}
\end{table} 

Eliminating the pole mass using \eqref{eq:massinmom} from the expression 
for the moment $M_k$ in terms of the moment $M_n$, we get  
\begin{equation} \label{MnRel} 
M_k=\left(\frac{9}{4}Q_Q^2\right)^{1-k/n}\left(M_n\right)^{k/n}\left(\frac{C_k}{(C_n)^{k/n}}\right) \ .
\end{equation}
Re-expanding in terms of $\alpha_s (m_Q)$ gives  
\begin{align}\label{eq:MkMnexp}
    M_k &= M_n^{k/n} \left(\frac{9}{4}Q_Q^2\right)^{1-k/n} a^{(0)}_{k,n} \nonumber \\
    & \times \left( 1 + \left(\frac{\alpha_s (\mu)}{4\pi}\right) a^{(1)}_{k,n} + \left(\frac{\alpha_s (\mu)} {4\pi}\right)^2 \left[a^{(1)}_{k,n} \beta_0 \ln \left(\frac{\mu^2}{m_Q^2} \right) +  a^{(2)}_{k,n} \right] + \cdots \right) .
\end{align}
Note that the logarithmic term ensures that this expression is in fact independent of $\mu$ to the order 
we calculate. The explicit expressions for the 
coefficients $a^{(i)}_{k,n}$ are given by:  
\begin{eqnarray}\label{eq:acoefs}
a^{(0)}_{k,n} &=&C_k^{(0)} (C_n^{(0)})^{-(k/n)}  \\
a^{(1)}_{k,n} &=& -\frac{C_n^{(1)}}{C_n^{(0)}}\frac{k}{n}+\frac{C_k^{(1)}}{C_k^{(0)}}\\
a^{(2)}_{k,n} &=&  \frac{(C_n^{(1)})^2}{(C_n^{(0)})^2} \frac{k^2+ k n}{2n^2} - \frac{C_k^{(1)} C_n^{(1)}}{C_k^{(0)}  C_n^{(0)}} \frac{k}{n} - \frac{C_n^{(2,0)}}{C_n^{(0)}} \frac{k}{n} + \frac{C_k^{(2,0)}}{C_k^{(0)}} 
\end{eqnarray}

We list the numerical values of the $C_i$ coefficients for $\mu=m_Q$ in Appendix~\ref{app:coeffs}. Using those, we obtain the numerical values for the $a_{k,n}^{(1)}$ and $a_{k,n}^{(2)}$ coefficients in Table~\ref{table:coeff}. For completeness, we give the $a_{k,n}^{(0)}$ coefficients in Appendix~\ref{app:coeffs}. We use $n_l=4$ which applies to the $B$ meson. However, we note that also for the charm-quark, with $n_l=4$, we observe a similar behavior.  

The values of $a^{(j)}_{k,n}$ strongly depend on the ration $k/n$. We consider inverse 
moments up to $n = 7$, which means that $1/7 \le k/n \le 7$. The re-expansion of the expression 
(\ref{MnRel}) in powers of $\alpha_s$ thus exhibits this strong $k/n$ dependence which jeopardizes 
the convergence of the resulting perturbative series. More generally, the expression (\ref{MnRel}) 
can be written as 
\begin{equation}
M_k = \Phi(M_n) 
\end{equation} 
and the re-expansion of $M_k$ involves the Taylor series of the function $\Phi$, and small values of 
$k/n$ correspond to small values of the derivatives of $\Phi$. 

In turn, this means that moments with $k/n \le 1$ should yield the most reliable results, and so we focus on the 
upper right triangle of tTable \ref{table:coeff}. Although we work only to $\alpha_s^2$, Table  \ref{table:coeff} shows that the convergence of the perturbative series relating $M_k$ with $M_n$ 
works best for $k \sim n$, since $a_{k,n}^{(1)}$ is of the same order as $a_{k,n}^{(2)}$. 

This can also be seen by computing the ``typical scale'' for which the ${\cal O} (\alpha_s^2)$ 
contribution vanishes. These scales are shown in Table \ref{table:scalesetting} in units of $m_Q$. 
For the upper-right triangle and for $k \sim n$ we find values of the order $m_Q$, which may be 
taken as an indication of convergence of the series expressed in terms of $\alpha_s (m_Q)$. 

In fact, as pointed out above, we do not expect a convergent series here, since the power corrections 
will eventually induce factorial divergences. However, in the case of the inverse moments of the 
$e^+ e^- \to $ hadrons cross section, the power corrections seem to be really small, such that 
the asymptotic behaviour induced by the power corrections is not visible in the perturbative series 
up to $\alpha_s^2$.

\begin{table}[t]
\centering
\begin{tabular}{cc|ccccccc}
&\multicolumn{8}{ c }{$n$} \\
\multirow{8}{*}{$k$}	&&	1	&	2	&	3	&	4	&	5	&	6	&	7	\\
\hline
 &1 & & 1.32 & 1.05 & 0.902 & 0.812 & 0.750 & 0.704 \\
  &2 & 2.10 &  & 0.811 & 0.670 & 0.584 & 0.526 & 0.484 \\
 &3&  2.61 & 1.18 &  & 0.637 & 0.537 & 0.472 & 0.425 \\
  &4 & 3.59 & 1.40 & 0.888 &  & 0.544 & 0.465 & 0.410 \\
 &5 &  5.23 & 1.74 & 1.03 & 0.737 &   & 0.484 & 0.418 \\
  &6 & 7.89 & 2.26 & 1.24 & 0.844 & 0.642 &    & 0.441 \\
 &7 &  12.2 & 3.00 & 1.52 & 0.991 & 0.729 & 0.576 &  \\
\end{tabular}
\caption{The scale $\mu/m_Q$ at which the $\alpha_s^2$-contribution in \eqref{eq:MkMnexp} vanishes for different combinations of $n$ and $k$ assuming $n_l=4$. }
\label{table:scalesetting}
\end{table} 

\subsection{Inclusive semileptonic decays of bottom hadrons}

The heavy quark expansion expresses inclusive $B$ decay rates in terms of a systematic 
expansion in inverse powers of the quark mass $m_b$. Like the inverse moments of the cross 
section for $e^+ e^- \to $ hadrons it is based on an OPE where the leading term is given 
by the perturbative (i.e. partonic) result. In particular, the perturbative result will 
depend on the $b$ quark mass. 

In full QCD, the heavy quark $Q$ has the equation of motion
\begin{equation} \label{DE}
(i \fmslash{D} - m_{\rm pole}) Q(x) = 0 
\end{equation} 
where $D_\mu$ is the usual QCD covariant derivative including the gluon fields 
and $m_{\rm pole}$ is the pole mass of the heavy quark, which we use as a starting point.  
It is in general defined as the pole of the perturbatively calculated quark propagator, 
but - since the quarks are not the asymptotic states of QCD - it cannot be assigned a 
physical meaning such as for the electron mass in QED.   
To this end, the pole mass remains a perturbative concept.  

In order to set up the HQE, we re-define the heavy quark field in (\ref{DE}) 
 \begin{equation} \label{GT} 
 Q(x) = \exp [-i m (vx)] Q_v (x) 
\end{equation}  
with some mass parameter $m$ which we will specify below. Inserting this, we find 
\begin{equation}
	(i \fmslash{D} - \fmslash{v} \delta m - (1- \fmslash{v}) m_{\rm pole} ) Q_\Lambda (x) = 0  
     \qquad \delta m =  m_{\rm pole} - m \, .
\end{equation}

The expansion in inverse powers of the quark mass is set up by assuming that 
$i D_\mu \sim \Lambda_{\rm QCD} \ll m_{\rm pole}$. In 
the following, we specify the mass parameter $m$ such that  
the ``residual mass term'' $\delta m$ is also counted like $ \delta m  \sim \Lambda_{\rm QCD}$. 
Introducing a new covariant derivative by 
\begin{equation} \label{newcov}
	i \nabla_\mu = i D_\mu - v_\mu \delta m 
\end{equation}
we can write 
\begin{equation}
	(i \fmslash{\nabla} + (\fmslash{v}-1) m_{\rm pole} ) Q_\Lambda (x) = 0 
\end{equation}
which means that the derivation of the heavy mass limit an the HQE proceeds in the usual way, 
except that all covariant derivatives are replaced by $D \to \nabla$. This leads to the quadratic form of the equation of motion:
\begin{equation}
	(i \fmslash{\nabla} + (\fmslash{v}+1) m_{\rm pole} )(i \fmslash{\nabla} + (\fmslash{v}-1) m_{\rm pole} ) Q_\Lambda (x) = [(i \fmslash{\nabla})^2 + 2 m_{\rm pole} (i v \nabla)] Q_v(x) =  0
\end{equation}
which in particular means 
\begin{equation}
 (i v \nabla) Q_v(x) =  {\cal O} (1/m) \quad \mbox{or} \quad  (i v D) Q_v(x) =  \delta m \, Q(x) + {\cal O} (1/m)
\end{equation} 

Turning to inclusive decays, the tree level contribution for inclusive semileptonic and radiative decays can be constructed from the external field propagator 
\begin{equation}
    \frac{1}{\fmslash{Q}+i \fmslash{D}-m_q} \quad \mbox{with} \quad Q = m v - q
\end{equation}
where $q$ is the momentum transferred to leptons of photons, $m$ is the chosen mass parameter 
of the field redefinition in (\ref{GT}) and $m_q$ is the mass of the final state quark. 
Expanding the propagator according to  \cite{Dassinger:2006md,Mannel:2010wj} gives
\begin{equation}
\frac{1}{\fmslash{Q}+i \fmslash{D}-m_q} = 
\frac{1}{\fmslash{Q} -m_q} - \frac{1}{\fmslash{Q} -m_q} (i \fmslash{D}) \frac{1}{\fmslash{Q} -m_q} + \cdots  
\end{equation}
We note that the relevant matrix element is constrained by the equation of motion, i.e. 
\begin{equation}
    \langle B (v) | \bar{b}_{v ,\alpha} iD_\mu b_{v, \beta} | B(v) \rangle = v_\mu    
      \langle B (v) | \bar{b}_{v ,\alpha} (ivD)  b_{v, \beta} | B(v) \rangle 
    = v_\mu \delta m  \langle B (v) | \bar{b}_{v, \alpha}  b_{v, \beta} | B(v) \rangle 
\end{equation}
where $\alpha,\beta$ are spinor indices. Inserting this yields 
\begin{equation}
\frac{1}{\fmslash{Q}+i \fmslash{D}-m_c} = 
\frac{1}{\fmslash{Q} -m_c} - \frac{1}{\fmslash{Q} -m} (\fmslash{v} \delta m) \frac{1}{\fmslash{Q} -m} + \cdots  
= \frac{1}{\fmslash{Q}+ \fmslash{v} \delta m  -m}
\end{equation}
which means that the pole mass gets re-installed in the denominator, since 
\begin{equation}
    \fmslash{Q}+ \fmslash{v} \delta m = (m+\delta m ) \fmslash{v} - \fmslash{q} = m_{\rm pole} \fmslash{v} - \fmslash{q}
\end{equation}
In general, the mass which appears in the denominator is the one which appears in the 
Lagrangain, i.e. in the equation of motion (\ref{DE}).

We consider now the leading term of the HQE, which is simply the partonic result. Computing 
QCD corrections in the pole scheme yields a perturbative series suffering from factorial divergences. 
It has been shown long ago  
\cite{Neubert:1994wq,Beneke:1994bc,Beneke:1994sw} that these divergences 
cancel against the ones induced by the pole mass, such that the renormalon at $u=1/2$ is cancelled, 
at least when using the bubble chain approximation. 

We assume that this remains true in full QCD and study the case where we use a renormalized 
short-distance mass for the field redefinition. Using (\ref{masrel}), we find 
\begin{equation}
 \delta m = m_Q^{\rm ren}  
    \sum_{i=1}^\infty r_i \left( \frac{\alpha_s  (m_Q^{\rm ren})}{\pi}
    \right)^i  
\end{equation} 
which starts at order $\alpha_s$. As argued above, these terms induced by $\delta m$ 
should cancel, at least, the divergences related to the $u=1/2$ renormalon of the 
perturbative series for the inclusive $b \to u$ rate computed in the pole scheme.

Now we turn to the strategy eliminating the mass in favour of an observable. We proceed in 
a similar way as in the case of the inverse moments of the $e^+e^- \to$ hadrons cross section 
and replace the quark mass. This could be done on the one hand by using spectral moments of 
inclusive semileptonic decays, but we shall proceed by making use of the inverse moments of the 
$e^+e^- \to$ hadrons cross section. We consider this to be the more interesting case, since this 
involves now very different observables measured by very different experiments.  

We consider a simple case, which is the total rate for the charmless inclusive semileptonic 
decay $B \to X_u \ell \bar{\nu}$. The leading term of the HQE for this process is the partonic result 
which reads to ${\cal O} (\alpha_s^2)$ in terms of the pole mass 
\begin{align} \label{eq:b2urate}
    \Gamma(B\to X_u \ell \bar{\nu}) &=  \frac{G_F|V_{ub}|^2m_{\rm pole}^5}{192 \pi^3} \nonumber \\ 
    & \times \left(1+\frac{\alpha_s}{\pi}b_1+\left(\frac{\alpha_s}{\pi}\right)^2 \left[b_2+\beta_0 b_1\ln \left(\frac{\mu^2}{m_Q^2}\right)\right] + \cdots \right)\ ,
\end{align}
where \cite{vanRitbergen:1999gs}
\begin{align}
b_1&= C_F\left(\frac{25}{8}-3\zeta_2\right)  \\ 
b_2&= C_AC_F\left[\frac{154927}{10368}-\frac{53}{2}\zeta_2\ln (2)+\frac{95}{27}\zeta_2-\frac{383}{72}\zeta_3+\frac{101}{16}\zeta_4 \right] \\ \nonumber 
&+ C_F^2\left[\frac{11047}{2592}+53\zeta_2\ln (2)-\frac{1030}{27}\zeta_2-\frac{223}{36}\zeta_3+\frac{67}{8}\zeta_4 \right] \\ \nonumber 
&+ C_FT_Fn_f\left[-\frac{1009}{288}+\frac{77}{36}\zeta_2+\frac{8}{3}\zeta_3\right] + C_FT_F \left[\frac{6335}{192}-\frac{9}{2}\zeta_2-24\zeta_3\right]  
\end{align}
Numerically, we have $b_1=-2.4, b_2=-21.3$ and $b_2/b_1=8.8$. 

We now replace the pole mass in \eqref{eq:b2urate} using \eqref{eq:massinmom} \footnote{The idea to replace the quark-mass in the inclusive rate for the $b\to c \ell \nu$ rate by moments of the $e^+e^-$ moments was already proposed in \cite{Penin:1998kx,Penin:1998wj}. Here they used a different approach for the $C_i$'s based on an expansion around the heavy quark threshold.}. Re-expanding in $\alpha_s$ gives: 
\begin{align}
  \Gamma(B\to X_u \ell \bar{\nu})
&\sim \left(\frac{C_n^{(0)}}{M_n}\right)^{5/(2n)} \nonumber \\
& \times \left(1+\frac{\alpha_s}{\pi}d_n^{(1)}+\left(\frac{\alpha_s}{\pi}\right)^2 \left[d_n^{(2)} + d_n^{(1)}\beta_0 \ln\left(\frac{\mu^2}{m_Q^2}\right) \right]+ \cdots \right),\label{inserted_gam}
\end{align}
with
\begin{align}\label{eq:dexp}
d_n^{(1)}&=\left(b_1+\frac{5}{2n}\frac{C_n^{(1)}}{C_n^{(0)}}\right)\nonumber \\
d_n^{(2)}&=\left(\frac{5}{2n}\frac{C_n^{(2,0)}}{C_n^{(0)}}+\frac{25}{8 n^2}\left(\frac{C_n^{(1)}}{C_n^{(0)}}\right)^2-\frac{5}{4n}\left(\frac{C_n^{(1)}}{C_n^{(0)}}\right)^2+\frac{5}{2n}b_1\frac{C_n^{(1)}}{C_n^{(0)}}+b_2\right) \, . 
\end{align}
Note that a factor $5/(2n)$ arises because of the $m_b^5$ dependence of the rate. This corresponds to 
the $k/n$ dependence of the inverse moments of the $e^+ e^- \to $ hadrons cross section discussed in 
the previous section. 

\begin{table}[t]
\centering
\begin{tabular}{c|ccccccc}
&\multicolumn{7}{ c }{$n$} \\
& 1	&	2	&	3	&	4	&	5	&	6	&	7	\\
\hline
 $d_n^{(1)}$ & 10.24 &  7.29 & 5.85 & 4.94 & 4.29 & 3.80 & 3.41\\
  $d_n^{(2)}$ & 70.41 &  49.45 & 39.69 & 33.70 & 29.52 & 26.40 & 23.93 \\\
    $d_n^{(2)}/d_n^{(1)}$ & 6.87 &  6.79 & 6.78 & 6.81 & 6.89 & 6.95 & 7.03\\\hline\hline
 $\mu/m_Q$ &  0.167 & 0.170 & 0.170 & 0.169 & 0.166 & 0.163  & 0.160 \\
\end{tabular}
\caption{The coefficients $d_n^{(1)}$ and $d_n^{(2)}$ defined in \eqref{eq:dexp} and the value $\mu/m_Q$ for which the $\alpha_s^2$ contribution vanishes.}  
\label{table:dval}
\end{table} 

We give the numerical values for the coefficients $d_{1,2}^{(n)}$ in table~\ref{table:dval}. 
We first note that the coefficients become smaller as $n$ increases. Furthermore, compared to the expressions in the pole 
scheme, the sign of the coefficients has changed, since both $b_1$ and $b_2$ are negative.  The ratio of the coefficients 
$d_2^{(n)}/d_1^{(n)}$ ranges between 6.87 and 7.03 (see Table/~\ref{table:dval}) and thus is 
not particularly small, indicating that the convergence of the pertrurbative series is not 
strongly improved, in particular once we compare to $b_2/b_1=8.8$, the values obtained in the 
pole scheme. 

Following the arguments given above this suggest the interpretation that 
removing the renormalons related to the mass 
does not significantly shift the onset of the asymptotic behaviour of the perturbative series 
relating the $B$-meson decay rate and the inverse moments. In order to improve this one would need 
to include renormalons of higher values of $u$, which then requires to include also the power 
corrections in the $B$ decay rate. This lies beyond the scope of our present paper. 

In Table~\ref{table:dval}, we also quote the scale at which the coefficient of the 
$\alpha_s^2$ corrections vanish. This turns out to be a scale of about 800 MeV and is 
remarkably constant for the various values of $n$.

\section{Conclusion}

It is known since almost thirty years that perturbative expansions in QCD cannot be 
disentangled from its non-perturbative features, since the quarks and gluons never appear 
as asymptotic states. The tool to disentangle perturbative from non-perturbative effects 
is the OPE, which yields, on top of the perturbative expansion, non-perturbative parameters, which are on the 
one hand the quark masses, and on the other hadronic matrix elements such as condensates
and HQE parameters. 

In heavy-quark physics, the precision of predictions heavily depends on the treatment 
of the heavy-quark mass. The pole mass, the usual starting point for the perturbative calculation, suffers from renormalon ambiguities that hinder a systematic expansion in this mass. Motivated by the assertion that these ambiguities cancel between the perturbative series and properly defined non-perturbative quantities, including the quark-masses, we discussed an alternative treatment of the heavy-quark mass by replacing it with 
physical observables. In this paper, we made a first numerical analysis of this idea. Assuming the pattern of cancellations of renormalon ambiguities as suggested 
by many seminal papers from the mid nineties mentioned previously, we used the known information on the perturbative 
series for various observables to study the behaviour of the resulting perturbative series
relating these observables.  

We found that, using the known perturbative results up to  $\alpha_s^2$, for the relation between different inverse moments of the $e^+ e^- \to$ hadrons 
cross section, that the perturbative series in fact improves significantly. We suggest that this 
correlates with the fact that the power correction for this observable only start at $1/m^4$. In addition, the hadronic matrix element, the gluon condensate is very small. However, we also found that using the same reasoning for the 
relation between the decay rate for $B \to X_u \ell \bar{\nu}$  and the inverse moments 
of the $e^+ e^- \to$ hadrons cross section does not significantly improve the perturbative 
series, which may be related to the presence of power corrections which start in this case at 
$1/m^2$. 

In view of the fact that the perturbative series will remain asymptotic, since we truncated 
the OPE and took only the leading term into account, we expect that including 
more observables will shift the onset of the divergent behaviour to even higher orders. Therefore, to push this idea further for the $B$ meson (or $D$ meson) decay rate, the power corrections have to be investigated, which means to include more observables in order to fix the unknown matrix elements. Furthermore, the method
to remove more and more renormalon ambiguities with higher values of $u$ may be refined by 
clever choices of observables, possibly giving us more confidence in the methods used in 
heavy-quark physics.  

Finally, this alternative strategy for the quark mass may also shed some light on the question, if the HQE can be used as a precision tool for charm decays. Removing the charm mass from the OPE expressions 
by inserting observables will eventually reveal, if the HQE, combined with the perturbative 
expansion in $\alpha_s (m_c)$, is a valid method to deal with charm decays.

\section*{Acknowledgements}
We thank A. Pivovarov for useful discussions. This research has been supported by the Deutsche Forschungsgemeinschaft 
(DFG, German Research Foundation) under grant 396021762 - TRR 257.

\appendix
\section{Coefficients $C_i$}\label{app:coeffs}
In this appendix, we list the numerical value for the coefficients $C_i$ entering in \eqref{eq:Cienter}. 
We first write 
\begin{equation}
C_n = C_n^{(0)} + \frac{\alpha_s}{\pi} C_F{C}_{F,n}^{(1)} + \left(\frac{\alpha_s}{\pi}\right)^2 \left[C_F^2 C_{A,n}^{(2)} + C_A C_F C_{NA,n}^{(2)}+ C_F T_f n_l C_{l,n}^{(2)} + C_F T_f C_{F,n}^{(2)}\right] \ ,
\end{equation}
where $C_A = 3, C_F = 4/3, T_f =1/2$. Here $n_l$ is the number of light massless fermions, which is related to the number of fermions $n_f$ via $n_l = n_f-1$. The different $C$ coefficients can be found in \cite{Chetyrkin:1997mb}, which also contain the logarithmic terms. Compared to \eqref{eq:Cienter}, we have $C_n^{(1)} \equiv C_F {C}_{F,n}^{(1)}$ and
\begin{equation}
C_n^{(2,0)}\equiv \left[C_F^2 C_{A,n}^{(2)} + C_A C_F C_{NA,n}^{(2)}+ C_F T_f n_l C_{l,n}^{(2)} + C_F T_f C_{F,n}^{(2)}\right]_{\mu = m_Q} . 
\end{equation}

Taking $\mu=m_Q$, the coefficients take the values given in Table~\ref{table:coeffsnum}. For the $b$ quark, we have $n_f=5$ and $n_l =4$, which we use for the numerical results in the main text.

For completeness, we give the coefficient $a_{k,n}^{(0)}$ given in \eqref{eq:acoefs}. Their numerical values are given in Table~\ref{table:a0coef}. 

\begin{table}[t]
\centering
\begin{tabular}{c|c|c|c|c|c|c}
$n$ & $C_{n}^{(0)}$ & ${C}_{F,n}^{(1)}$ & $C_{A,n}^{(2)}$ & $C_{NA,n}^{(2)}$ & $C_{l,n}^{(2)}$  & $C_{F,n}^{(2)}$ \\
\hline\hline
1 & 1.06667 & 4.04938 & 5.07543 & 7.09759 & -2.33896 &  0.72704 \\
2 & 0.45714 & 2.66074 & 6.39333 & 6.31108 & -2.17395 & 0.26711\\
3 & 0.27090 &  2.01494 & 6.68902 & 5.39768 & -1.89566 &  0.14989\\
4 & 0.18470 & 1.62997 & 6.68456 & 4.69907 & -1.67089 &  0.09947\\
5 & 0.13640 & 1.37194 & 6.57434 & 4.16490 & -1.49436 & 0.07230\\
6 & 0.10609 & 1.18616 & 6.42606 & 3.74591 & -1.35348 & 0.05566\\
7 & 0.08558 & 1.04568 & 6.26672 & 3.40886 & -1.23871 & 0.04459\\
\end{tabular}
\caption{Numerical values for the coefficients $C_n^{(i)}$ from \cite{Chetyrkin:1997mb} at $\mu=m_Q$.}
\label{table:coeffsnum}
\end{table}

\begin{table}[t]
\centering
\begin{tabular}{cc|ccccccc}
&\multicolumn{8}{ c }{$n$} \\
\multirow{8}{*}{$k$}	&&	1	&	2	&	3	&	4	&	5	&	6	&	7	\\
\hline
 &1 &  1.00 & 1.58 & 1.65 & 1.63 & 1.59 & 1.55 & 1.52 \\
  &2 & 0.40 & 1.00 & 1.09 & 1.06 & 1.01 & 0.97 & 0.92 \\
 &3&  0.22 & 0.88 & 1.00 & 0.96 & 0.90 & 0.83 & 0.78 \\
  &4 &  0.14 & 0.88 & 1.05 & 1.00 & 0.909 & 0.82 & 0.75 \\
 &5 &   0.10 & 0.97 & 1.20 & 1.13 & 1.00 & 0.89 & 0.79 \\
  &6 &  0.07 & 1.11 & 1.45 & 1.34 & 1.16 & 1.00 & 0.87 \\
 &7 &  0.06 & 1.32 & 1.80 & 1.64 & 1.39 & 1.17 & 1.00 \\
\end{tabular}
\caption{The $a_{k,n}^{(0)}$ coefficients defined in \eqref{eq:acoefs}. }
\label{table:a0coef}
\end{table}

\bibliographystyle{jhep} 
\bibliography{refs.bib} 

\providecommand{\href}[2]{#2}\begingroup\raggedright\begin{thebibliography}{10}

\bibitem{Beneke:1998ui}
M.~Beneke, \emph{{Renormalons}},
  \href{http://dx.doi.org/10.1016/S0370-1573(98)00130-6}{\emph{Phys. Rept.}
  {\bf 317} (1999) 1--142}, [\href{http://arxiv.org/abs/hep-ph/9807443}{{\tt
  hep-ph/9807443}}].

\bibitem{Bigi:1994em}
I.~I.~Y. Bigi, M.~A. Shifman, N.~G. Uraltsev and A.~I. Vainshtein, \emph{{The
  Pole mass of the heavy quark. Perturbation theory and beyond}},
  \href{http://dx.doi.org/10.1103/PhysRevD.50.2234}{\emph{Phys. Rev. D} {\bf
  50} (1994) 2234--2246}, [\href{http://arxiv.org/abs/hep-ph/9402360}{{\tt
  hep-ph/9402360}}].

\bibitem{Beneke:1994sw}
M.~Beneke and V.~M. Braun, \emph{{Heavy quark effective theory beyond
  perturbation theory: Renormalons, the pole mass and the residual mass term}},
  \href{http://dx.doi.org/10.1016/0550-3213(94)90314-X}{\emph{Nucl. Phys. B}
  {\bf 426} (1994) 301--343}, [\href{http://arxiv.org/abs/hep-ph/9402364}{{\tt
  hep-ph/9402364}}].

\bibitem{Neubert:1994wq}
M.~Neubert and C.~T. Sachrajda, \emph{{Cancellation of renormalon ambiguities
  in the heavy quark effective theory}},
  \href{http://dx.doi.org/10.1016/0550-3213(95)00032-N}{\emph{Nucl. Phys. B}
  {\bf 438} (1995) 235--260}, [\href{http://arxiv.org/abs/hep-ph/9407394}{{\tt
  hep-ph/9407394}}].

\bibitem{Chetyrkin:2010ic}
K.~Chetyrkin, J.~H. Kuhn, A.~Maier, P.~Maierhofer, P.~Marquard, M.~Steinhauser
  et~al., \emph{{Precise Charm- and Bottom-Quark Masses: Theoretical and
  Experimental Uncertainties}},
  \href{http://dx.doi.org/10.1007/s11232-012-0024-7}{\emph{Theor. Math. Phys.}
  {\bf 170} (2012) 217--228}, [\href{http://arxiv.org/abs/1010.6157}{{\tt
  1010.6157}}].

\bibitem{Bigi:1994ga}
I.~I.~Y. Bigi, M.~A. Shifman, N.~G. Uraltsev and A.~I. Vainshtein, \emph{{Sum
  rules for heavy flavor transitions in the SV limit}},
  \href{http://dx.doi.org/10.1103/PhysRevD.52.196}{\emph{Phys. Rev. D} {\bf 52}
  (1995) 196--235}, [\href{http://arxiv.org/abs/hep-ph/9405410}{{\tt
  hep-ph/9405410}}].

\bibitem{Fael:2020iea}
M.~Fael, K.~Sch\"onwald and M.~Steinhauser, \emph{{Kinetic Heavy Quark Mass to
  Three Loops}},
  \href{http://dx.doi.org/10.1103/PhysRevLett.125.052003}{\emph{Phys. Rev.
  Lett.} {\bf 125} (2020) 052003}, [\href{http://arxiv.org/abs/2005.06487}{{\tt
  2005.06487}}].

\bibitem{Fael:2020njb}
M.~Fael, K.~Sch\"onwald and M.~Steinhauser, \emph{{Relation between the
  $\overline{\mathrm{MS}}$ and the kinetic mass of heavy quarks}},
  \href{http://dx.doi.org/10.1103/PhysRevD.103.014005}{\emph{Phys. Rev. D} {\bf
  103} (2021) 014005}, [\href{http://arxiv.org/abs/2011.11655}{{\tt
  2011.11655}}].

\bibitem{Hoang:1998ng}
A.~H. Hoang, Z.~Ligeti and A.~V. Manohar, \emph{{B decay and the Upsilon
  mass}}, \href{http://dx.doi.org/10.1103/PhysRevLett.82.277}{\emph{Phys. Rev.
  Lett.} {\bf 82} (1999) 277--280},
  [\href{http://arxiv.org/abs/hep-ph/9809423}{{\tt hep-ph/9809423}}].

\bibitem{Hoang:1998hm}
A.~H. Hoang, Z.~Ligeti and A.~V. Manohar, \emph{{B decays in the upsilon
  expansion}}, \href{http://dx.doi.org/10.1103/PhysRevD.59.074017}{\emph{Phys.
  Rev. D} {\bf 59} (1999) 074017},
  [\href{http://arxiv.org/abs/hep-ph/9811239}{{\tt hep-ph/9811239}}].

\bibitem{Gambino:2010jz}
P.~Gambino and J.~F. Kamenik, \emph{{Lepton energy moments in semileptonic
  charm decays}},
  \href{http://dx.doi.org/10.1016/j.nuclphysb.2010.07.019}{\emph{Nucl. Phys. B}
  {\bf 840} (2010) 424--437}, [\href{http://arxiv.org/abs/1004.0114}{{\tt
  1004.0114}}].

\bibitem{Fael:2019umf}
M.~Fael, T.~Mannel and K.~K. Vos, \emph{{The Heavy Quark Expansion for
  Inclusive Semileptonic Charm Decays Revisited}},
  \href{http://dx.doi.org/10.1007/JHEP12(2019)067}{\emph{JHEP} {\bf 12} (2019)
  067}, [\href{http://arxiv.org/abs/1910.05234}{{\tt 1910.05234}}].

\bibitem{Martinelli:1995zw}
G.~Martinelli, M.~Neubert and C.~T. Sachrajda, \emph{{The Invisible
  renormalon}},
  \href{http://dx.doi.org/10.1016/0550-3213(95)00613-3}{\emph{Nucl. Phys. B}
  {\bf 461} (1996) 238--258}, [\href{http://arxiv.org/abs/hep-ph/9504217}{{\tt
  hep-ph/9504217}}].

\bibitem{Martinelli:1996pk}
G.~Martinelli and C.~T. Sachrajda, \emph{{On the difficulty of computing higher
  twist corrections}},
  \href{http://dx.doi.org/10.1016/0550-3213(96)00415-4}{\emph{Nucl. Phys. B}
  {\bf 478} (1996) 660--686}, [\href{http://arxiv.org/abs/hep-ph/9605336}{{\tt
  hep-ph/9605336}}].

\bibitem{Chetyrkin:1996cf}
K.~G. Chetyrkin, J.~H. Kuhn and M.~Steinhauser, \emph{{Three loop polarization
  function and O($\alpha_s^2$) corrections to the production of heavy quarks}},
  \href{http://dx.doi.org/10.1016/S0550-3213(96)00534-2}{\emph{Nucl. Phys. B}
  {\bf 482} (1996) 213--240}, [\href{http://arxiv.org/abs/hep-ph/9606230}{{\tt
  hep-ph/9606230}}].

\bibitem{Chetyrkin:1997mb}
K.~G. Chetyrkin, J.~H. Kuhn and M.~Steinhauser, \emph{{Heavy quark current
  correlators to O($\alpha_s^2$)}},
  \href{http://dx.doi.org/10.1016/S0550-3213(97)00481-1}{\emph{Nucl. Phys. B}
  {\bf 505} (1997) 40--64}, [\href{http://arxiv.org/abs/hep-ph/9705254}{{\tt
  hep-ph/9705254}}].

\bibitem{Boughezal:2006px}
R.~Boughezal, M.~Czakon and T.~Schutzmeier, \emph{{Charm and bottom quark
  masses from perturbative QCD}},
  \href{http://dx.doi.org/10.1103/PhysRevD.74.074006}{\emph{Phys. Rev. D} {\bf
  74} (2006) 074006}, [\href{http://arxiv.org/abs/hep-ph/0605023}{{\tt
  hep-ph/0605023}}].

\bibitem{Chetyrkin:2006xg}
K.~G. Chetyrkin, J.~H. Kuhn and C.~Sturm, \emph{{Four-loop moments of the heavy
  quark vacuum polarization function in perturbative QCD}},
  \href{http://dx.doi.org/10.1140/epjc/s2006-02610-y}{\emph{Eur. Phys. J. C}
  {\bf 48} (2006) 107--110}, [\href{http://arxiv.org/abs/hep-ph/0604234}{{\tt
  hep-ph/0604234}}].

\bibitem{Kuhn:2007vp}
J.~H. Kuhn, M.~Steinhauser and C.~Sturm, \emph{{Heavy Quark Masses from Sum
  Rules in Four-Loop Approximation}},
  \href{http://dx.doi.org/10.1016/j.nuclphysb.2007.04.036}{\emph{Nucl. Phys. B}
  {\bf 778} (2007) 192--215}, [\href{http://arxiv.org/abs/hep-ph/0702103}{{\tt
  hep-ph/0702103}}].

\bibitem{Shifman:1978bx}
M.~A. Shifman, A.~I. Vainshtein and V.~I. Zakharov, \emph{{QCD and Resonance
  Physics. Theoretical Foundations}},
  \href{http://dx.doi.org/10.1016/0550-3213(79)90022-1}{\emph{Nucl. Phys. B}
  {\bf 147} (1979) 385--447}.

\bibitem{Shifman:1978by}
M.~A. Shifman, A.~I. Vainshtein and V.~I. Zakharov, \emph{{QCD and Resonance
  Physics: Applications}},
  \href{http://dx.doi.org/10.1016/0550-3213(79)90023-3}{\emph{Nucl. Phys. B}
  {\bf 147} (1979) 448--518}.

\bibitem{Broadhurst:1994qj}
D.~J. Broadhurst, P.~A. Baikov, V.~A. Ilyin, J.~Fleischer, O.~V. Tarasov and
  V.~A. Smirnov, \emph{{Two loop gluon condensate contributions to heavy quark
  current correlators: Exact results and approximations}},
  \href{http://dx.doi.org/10.1016/0370-2693(94)90524-X}{\emph{Phys. Lett. B}
  {\bf 329} (1994) 103--110}, [\href{http://arxiv.org/abs/hep-ph/9403274}{{\tt
  hep-ph/9403274}}].

\bibitem{Colangelo:2000dp}
P.~Colangelo and A.~Khodjamirian, \emph{{QCD sum rules, a modern perspective}},
   \href{http://arxiv.org/abs/hep-ph/0010175}{{\tt hep-ph/0010175}}.

\bibitem{Dassinger:2006md}
B.~M. Dassinger, T.~Mannel and S.~Turczyk, \emph{{Inclusive semi-leptonic B
  decays to order $1/m_b^4$}},
  \href{http://dx.doi.org/10.1088/1126-6708/2007/03/087}{\emph{JHEP} {\bf 03}
  (2007) 087}, [\href{http://arxiv.org/abs/hep-ph/0611168}{{\tt
  hep-ph/0611168}}].

\bibitem{Mannel:2010wj}
T.~Mannel, S.~Turczyk and N.~Uraltsev, \emph{{Higher Order Power Corrections in
  Inclusive B Decays}},
  \href{http://dx.doi.org/10.1007/JHEP11(2010)109}{\emph{JHEP} {\bf 11} (2010)
  109}, [\href{http://arxiv.org/abs/1009.4622}{{\tt 1009.4622}}].

\bibitem{Beneke:1994bc}
M.~Beneke, V.~M. Braun and V.~I. Zakharov, \emph{{Bloch-Nordsieck cancellations
  beyond logarithms in heavy particle decays}},
  \href{http://dx.doi.org/10.1103/PhysRevLett.73.3058}{\emph{Phys. Rev. Lett.}
  {\bf 73} (1994) 3058--3061}, [\href{http://arxiv.org/abs/hep-ph/9405304}{{\tt
  hep-ph/9405304}}].

\bibitem{vanRitbergen:1999gs}
T.~van Ritbergen, \emph{{The Second order QCD contribution to the semileptonic
  $b\to u$ decay rate}},
  \href{http://dx.doi.org/10.1016/S0370-2693(99)00407-4}{\emph{Phys. Lett. B}
  {\bf 454} (1999) 353--358}, [\href{http://arxiv.org/abs/hep-ph/9903226}{{\tt
  hep-ph/9903226}}].

\bibitem{Penin:1998kx}
A.~A. Penin and A.~A. Pivovarov, \emph{{Bottom quark pole mass and $|V_{cb}|$
  matrix element from $R(e^+ e^- \to b \bar{b})$ and $\Gamma_{\rm sl}(b \to
  c\ell \nu_\ell)$ in the next to next-to-leading order}},
  \href{http://dx.doi.org/10.1016/S0550-3213(99)00182-0}{\emph{Nucl. Phys. B}
  {\bf 549} (1999) 217--241}, [\href{http://arxiv.org/abs/hep-ph/9807421}{{\tt
  hep-ph/9807421}}].

\bibitem{Penin:1998wj}
A.~A. Penin and A.~A. Pivovarov, \emph{{Next-to-next-to-leading order relation
  between $R(e^+ e^- \to b \bar{b})$ and $\Gamma_{\rm sl}(b\to l \nu_l)$ and
  precise determination of $|V_{cb}|$}},
  \href{http://dx.doi.org/10.1016/S0370-2693(98)01323-9}{\emph{Phys. Lett. B}
  {\bf 443} (1998) 264--268}, [\href{http://arxiv.org/abs/hep-ph/9805344}{{\tt
  hep-ph/9805344}}].

\end{thebibliography}\endgroup

\end{document}